\documentclass[journal]{IEEEtran}

\ifCLASSINFOpdf

\else

\fi

\newcommand\notype[1]{\unskip}

\renewcommand{\footnoterule}{%
	\kern 2pt
	\hrule width 5cm height 0.5pt
	\kern 3pt
}

\usepackage[mathscr]{euscript}

\DeclareSymbolFont{rsfs}{U}{rsfs}{m}{n}
\DeclareSymbolFontAlphabet{\mathscrsfs}{rsfs}

\makeatletter
\newcommand{\mathleft}{\@fleqntrue\@mathmargin0pt}

\usepackage{algorithmic}               
\usepackage{algorithm}  

\usepackage{blkarray, bigstrut}
\usepackage[utf8x]{inputenc}
\usepackage{amsmath}
\usepackage{amsthm}
\usepackage{graphicx}
\usepackage{tipa}
\usepackage{changes}
\usepackage{mathtools}
\usepackage{amssymb}

\ifCLASSOPTIONcompsoc
\usepackage[caption=false, font=normalsize, labelfont=sf, textfont=sf]{subfig}
\else
\usepackage[caption=false, font=footnotesize]{subfig}
\fi
\usepackage{tikz}
\usetikzlibrary{calc}
\newtheorem{remark}{Remark}

\newtheorem{theorem}{Theorem}

\begin{document}
	\title{PHY-Fed: An Information-Theoretic Secure Aggregation in Federated Learning in Wireless Communications}
	\author{Mitra Hassani,~\IEEEmembership{Student Member,~IEEE},~ Reza Gholizadeh,~\IEEEmembership{Student Member,~IEEE}}

	\markboth{}%
	{Shell \MakeLowercase{\textit{et al.}}: Bare Demo of IEEEtran.cls for IEEE Journals}
	\maketitle
	
	\begin{abstract}
Federated learning (FL) is a type of distributed machine learning at the
wireless edge which preserves the privacy of clients' data from adversaries and even the central server. Existing federated learning approaches
either use (i) secure multiparty computation (SMC) which is vulnerable
to inference or (ii) differential privacy which may decrease the test accuracy given a large number of parties with relatively small amounts of
data each. To tackle the problem with the existing methods in the literature, In this paper, we introduce PHY-Fed, a new framework that secures federated algorithms from information-theoretic point of view.
	\end{abstract}

	\begin{IEEEkeywords}
		Federated learning, Privacy, Physical layer security, Channel reciprocity, Cloud computing.
	\end{IEEEkeywords}

	\IEEEpeerreviewmaketitle
	\section{introduction} \label{intro}
In traditional machine learning (ML) schemes, data is centrally trained by one organization running the learning algorithm.	
Federated Learning (FL) is a recent technique in distributed machine learning (ML) that trains an algorithm locally at decentralized clients. This method has two major benefits, namely (i) accessing to larger data provided by different clients, and (ii) utilizing the computing power of all the clients to train a general model \cite{8870236,9076343,9042352,9014530,8952884,hamidi2019systems,mcmahan2017communication,hamidi2019systems,9833972}. 
In addition, in FL the clients collaboratively train models without the sharing of raw data. This is because clients often cannot share data due to some restrictions or competition between them.
To satisfy this condition, clients only exchange model parameters in lieu of the raw information. Training will be completed when the model parameters converge after enough number of iterations. At each iteration, the updated parameters calculated by the clients will be transmitted to a central server, called Parameter Server (PS) hereafter, for aggregation \cite{mcmahan2017communication}. 
The PS is assumed to be honest-but-curious, i.e., it is curious in interpreting the data of
individuals, however, it is honest in operations. The privacy in FL needs to meet two conditions:

\noindent $\bullet$~\textbf{Condition I}: the server remains oblivious to the
messages of all the clients, while it can interpret/use the aggregated messages.

\noindent $\bullet$~\textbf{Condition II}: each client is oblivious to the messages sent by the other clients, and therefore maintaining the privacy of
clients’ data from being eavesdropped by malicious adversaries. 

Two commonly methods used in FL that satisfy Conditions I and II are \textit{Cryptographic methods} and \textit{Differential privacy}. In the following, these two methods and their associated problems are briefly discussed, and the proposed approach for data aggregation presented in this paper is introduced. 

 \textit{Cryptographic methods:} In this type of technique, different users encrypt their local messages prior to sending it to the PS which ensures that the PS does not know the data on each device. Then, the PS operates on the encrypted messages, and finally decrypts the aggregated model to get the final result. There are two commonly used methods in this category, namely homomorphic encryption \cite{aono2017privacy} and  secure multiparty computation (SMC) \cite{bonawitz2017practical}. Nevertheless, there are two major issues with this type of secrecy:
 \begin{itemize}
     \item (i) Most commonly used encryption methods rely on the computational hardness of some mathematical problems. Nevertheless, due to continual advances in computer technology, and discovery of new computational techniques such as Quantum computing, eavesdroppers are becoming more equipped and intelligent, and thus long term effectiveness of such traditional techniques is questionable.
     \item {(ii) To preserve the privacy, such systems suffer from the exceedingly high computation overhead (otherwise breaking such schemes would become easy for adversaries).}
 \end{itemize} 

\textit{Differential privacy:} While satisfying the conditions in (), this technique guarantees that one single update does not have much influence on the output of PS model \cite{dwork2014algorithmic}. To be more specific, the inclusion of a single instance in the training dataset makes only negligible changes to the algorithm’s final output. Also, by means of injecting artificial noises to the uploaded parameters by the users, this method provides the model with protection against the inference attack. 
However, it leads to low accuracy when there is a large number of clients with relatively small amounts of data each.

To overcome these challenges, in this paper we propose a secure data aggregation algorithm exploiting Physical Layer Security (PLS) for FL models. Note that both \textit{Cryptographic methods} and \textit{Differential privacy}---which are applied at the third or higher layers of communication protocols---are independent from our algorithm, and therefore they could be integrated with our proposed method. Most PLS schemes are based on the idea of utilizing intrinsic randomness in wireless channels. This randomness is used to establish a common value between two legitimate nodes. As this randomness only depends on the characteristics of the channel between the two parties, an adversary will be oblivious to this randomness. It is shown that this randomness could be exploited to secure wireless communications from an information-theoretic perspective. 

To obtain a common random value by both parties, channel reciprocity is one of the main principles used in this realm. This feature implies that the transmitted signal from nodes $A$ and $B$ experiences almost the same fading in the links $A\rightarrow B$ and $B\rightarrow A$. Therefore, the phase change exerted over signals transmitted in the links $A\rightarrow B$ and $B\rightarrow A$ are the same. 
\section{System Model} \label{model}
ML algorithms often entail minimization of the empirical loss function of the form $F(\boldsymbol{\theta})=\frac{1}{D}\sum_{i=1}^D f(\boldsymbol{\theta},\boldsymbol{u}_i )$, where $\boldsymbol{\theta} \in \mathbb{C}^d$ are model parameters to be optimized, $\boldsymbol{u}_i$ for $1\leq i \leq D$ are the training data samples, and $f(.)$ is the loss function that depends on the ML model. Iterative Stochastic Gradient Descent (SGD) is often adopted to minimize the function $F(\boldsymbol{\theta})$, where the model parameters at iteration $t$, denoted by $\boldsymbol{\theta}^t$, are updated as $\boldsymbol{\theta}^{t+1}=\boldsymbol{\theta}^{t}-\eta^t \boldsymbol{g}(\boldsymbol{\theta})$ while satisfying $\mathbb{E}\{\boldsymbol{g}(\boldsymbol{\theta})\}=\nabla F(\boldsymbol{\theta})$, where $\eta^t$ is the learning rate. On the other hand, in distributed SGD (DSGD), clients train data samples in parallel while keeping a globally consistent parameter vector $\boldsymbol{\theta}^t$. To shed more light, in each iteration, client $s$ computes a gradient vector based on the global parameter vector with respect
to its local dataset, and sends back the result to the PS, so that PS updates
the global parameter vector as follows
\begin{align} \label{Eq:main_update}
\boldsymbol{\theta}^{t+1}= \boldsymbol{\theta}^t-\eta^t \frac{1}{S} \sum_{s=1}^S \boldsymbol{g}_s(\boldsymbol{\theta}^t),   
\end{align}
where $S$ is the number of clients; and also, $\boldsymbol{g}_s(\boldsymbol{\theta}^t)\triangleq \frac{1}{|\chi_s|} \sum_{\boldsymbol{u}_i \in \chi_s} \nabla f(\boldsymbol{\theta^t},\boldsymbol{u}_i)$ is the gradient
estimate of Client $s$ obtained from the global parameter $\boldsymbol{\theta}^t$ and its local dataset $\chi_s$. Iterations are carried out until a certain convergence criterion is met.

As seen, parallelism enjoys exploiting the computing power of $S$ workers, however, it brings up the issue of preserving privacy of the clients. Equation \eqref{Eq:main_update} shows that PS requires only the gradient estimate of Client $s$, i.e., $\boldsymbol{g}_s(\boldsymbol{\theta}^t)$, to update the global parameter for the model. Although $\boldsymbol{g}_s(\boldsymbol{\theta}^t)$ contains less information than the $C_s$'s raw data, PS can use only the updated parameters to infer some information about $\chi_s$ \cite{bonawitz2016practical}. Thus, Client's gradient estimates must be securely aggregated at the PS such that Conditions I and II are satisfied. In the following, by using the phase reciprocity of wireless channels, a secure algorithm for aggregating $\boldsymbol{g}_s(\boldsymbol{\theta}^t)$ at PS is presented. First, in Section \ref{Sec:recip}, the phase reciprocity and its masking capability is explained; then in Section \ref{Sec:recip} the algorithm is elaborated.

\section{Masking the Content of Gradient Vector}\label{Sec:recip}
In wireless channels, the distribution of the phase is uniform in $[0,2\pi)$ \cite{goldsmith2005wireless}.
Consider two clients $C_i$ and $C_j$ in the distributed FL model. We denote the channel phase exerted in the links $C_i\rightarrow C_j$ and $C_j\rightarrow C_i$ by $\phi_{ij}$ and $\phi_{ji}$, respectively. Then, if $C_k$ is another clients whose position is $\lambda/2$ aprart from $C_j$, then $\phi_{ik}$ and $\phi_{ij}$ are independent, where $\lambda$ is the radio wavelength. Therefore, in our case, all the $\phi_{ij}$ for $i\neq j$ are independent and identically distributed (i.i.d.) in $[0,2\pi)$. We use this property to design an information-theoretic secure algorithm for FL.

As discussed in Section \ref{model}, at each iteration, clients need to send their updated gradient vector to PS. For this purpose, we assume that the clients modulate their local gradient updates by using QPSK modulation (they can use any other type of PSK modulation, but for the sake of simplicity, we use QPSK). Therefore, the gradient vector must be first quantized to a desired level of accuracy, and converted to a binary string. Assume, this results in representing $\boldsymbol{g}_s(\boldsymbol{\theta}^t)$ by $L$ bits. In addition, to combat the effect of channel noise and other imperfections typical in wireless transmission, the $L$ bits representing the gradient vector should undergo FEC, adding $r$ bits of channel coding redundancy to the original $L$ bits. Therefore, $L+r=2m$ QPSK symbols are required to send the gradient vector to PS. We denote the modulated gradient vector by $\bold{MOD}({g}_s(\boldsymbol{\theta}^t))=\vec{\nu}_s^t$.

In the following, we elaborate on how the content of QPSK symbols could be completely masked.

First, consider the following theorem proved in \cite{hamidi2021secure}.
\begin{theorem} \label{theorem_angle}
Assume that angle $x$ is uniformly distributed in $[0, 2\pi)$, and  $y$ is another random angle, independent of $x$, distributed in  $[0, 2\pi)$ with certain Probability Distribution Function (PDF). Then modulo-$2\pi$ addition of $x$ and $y$ is also uniformly distributed in $[0, 2\pi)$.
\end{theorem}
Based on Theorem 1, if a client rotates QPSK constellation point $X$ by a random variable $\Phi$ uniformly distributed in $[0, 2\pi)$, the resulted QPSK symbol will be also uniformly distributed in $[0, 2\pi)$, and therefore carrying no information about its summands. Note that rotating in this manner is equivalent to modulo-$2\pi$ addition of $X$ and $\Phi$:
$Y=X\oplus\Phi$. Any other clients who observes $Y$, cannot extract any information about $X$, leading to $I(X;Y) = H(Y)-H(Y|X)=H(Y)-H(\Phi)=0$. Based on this observation and phase reciprocity in wireless channels, in the following Section, we will present a secure algorithm for FL models where in the privacy requirements are satisfied.

\begin{figure}[t!]
		\centering
		\includegraphics[width=0.8\columnwidth, trim = {0 0.0cm 0 0.0cm}]{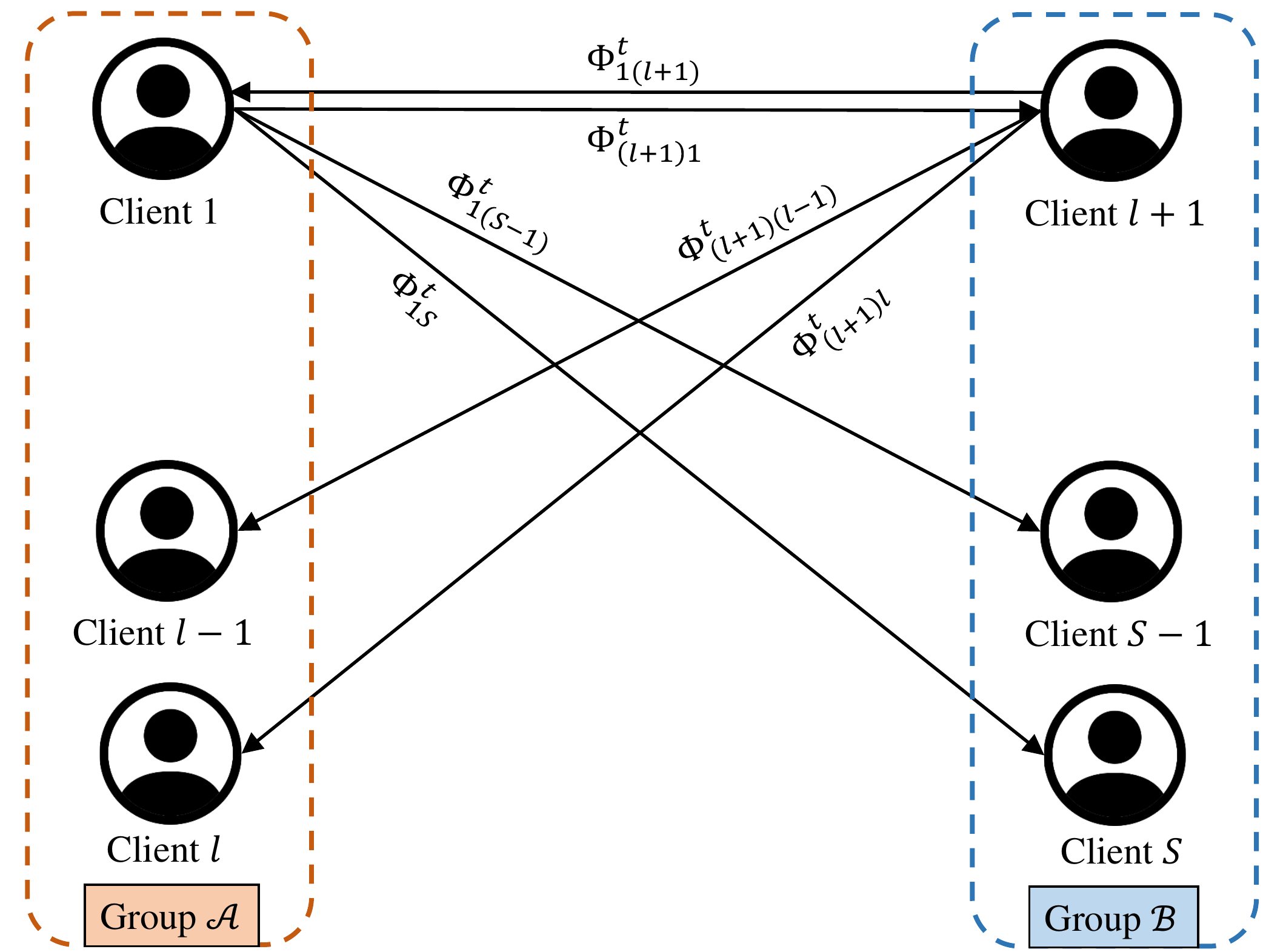}
	\caption{This figure shows how clients 1 and $l+1$, as two candidates from groups $\mathscr{A}$ and $\mathscr{B}$, find $\Phi_1^t$ and $\Phi_{l+1}^t$, respectively.
	Client 1 calculates $\Phi_1^t=\sum_{i=1}^{l} \Phi_{1i}^t$, and client $l+1$ finds $\Phi_{l+1}^t=\sum_{i=l+1}^{S} \Phi_{(l+1)i}^t$}.
	\label{fig:shared_phase}
\end{figure}

\begin{figure*} \label{Fig:alg}
  \includegraphics[width=\textwidth]{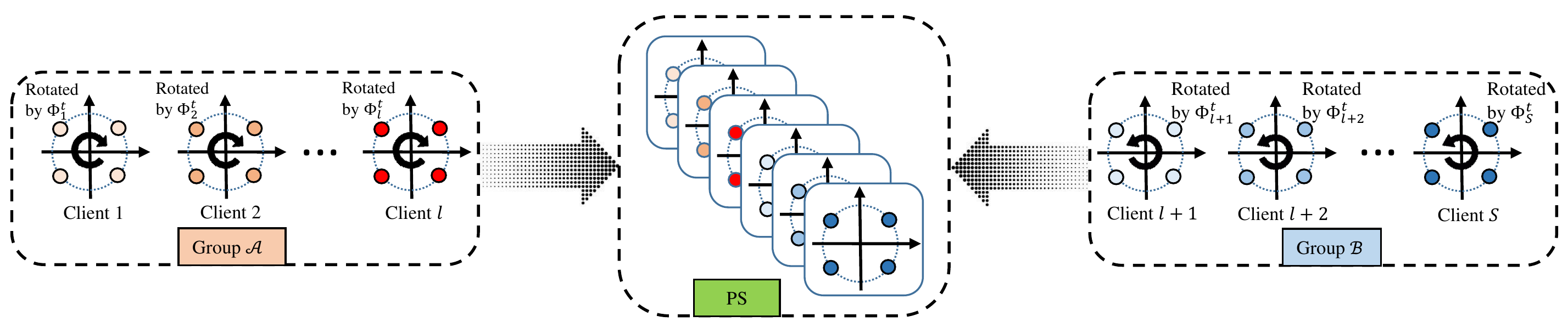}
  \caption{At each iteration $t$, the clients are divided into two groups, namely $\mathscr{A}$ and $\mathscr{B}$. The clients in group $\mathscr{A}$ rotates their QPSK constellations clockwise by their respective secret phases. On the other hand, the clients in group $\mathscr{B}$ rotates their QPSK constellations counter-clockwise by their respective secret phases. Once the data from both groups are aggregated at PS, the random phases added by the clients are removed.}
\end{figure*}

\section{A Secure Algorithm for Aggregation in FL}
In the first phase of our proposed algorithm for data aggregation in PS, the clients are randomly divided into two groups, namely group $\mathscr{A}$ and $\mathscr{B}$, such that in each group there is at least two clients. We use $\mathcal{G}(C_i)=\mathscr{A}$ to indicate that $C_i$ is a member of group $\mathscr{A}$. At each iteration $t$, clients in $\mathscr{A}$, mask the content of $\bold{MOD}({g}_s(\boldsymbol{\theta}^t))=\vec{\nu}_s^t$ by rotating the constellations counter clock-wise with the phase $\Phi_i^t=\sum_{j \notin \mathcal{G}(C_i)} \phi_{ij}^{t}$ to find $\mathscr{E}^{\oplus}(\vec{\nu}_i^t)=\vec{\nu}_i^t \oplus \Phi_i^t$ ($\Phi_i^t$ is equal to the sum of phases of the channels between a client and the clients in the other group, see Fig. \ref{fig:shared_phase}). On the other hand, clients in $\mathscr{B}$, rotate their QPSK constellation points clock-wise by $\Phi_i^t=\sum_{j \notin \mathcal{G}(C_i)} \phi_{ij}^{t}$, to find $\mathscr{E}^{\ominus}(\vec{\nu}_i^t)=\vec{\nu}_i^t \ominus \Phi_i^t$. Then all clients send their encrypted messages to PS., i.e., client $i$ in group $\mathscr{A}$ transmits $\mathscr{E}^{\oplus}(\vec{\nu}_i^t)$ and client $j$ in group $\mathscr{B}$ transmits $\mathscr{E}^{\ominus}(\vec{\nu}_j^t)$.

The Algorithm 1 explains the procedure systematically.

\begin{algorithm} \label{alg:01}
 \caption{Two-group structure}
 \begin{algorithmic}[1]
 \renewcommand{\algorithmicrequire}{\textbf{Input:}}
 \renewcommand{\algorithmicensure}{\textbf{Output:}}
 \REQUIRE The number of clients, namely $S$
 \STATE Each client generates random variable  
 \STATE Randomly divide the clients into two groups, namely $\mathscr{A}$ and $\mathscr{B}$ \label{alg:grouping};
 \IF{there is a group with less that two clients}
 \STATE{Return to Step \ref{alg:grouping}}
 \ELSE
 \STATE Label the clients in the first group by $\{1,2,...,l\}$, and those in the second group by $\{l+1,l+2,...,S\}$;
 \FOR{$i$ = 1,...,$S$ in parallel do}
 \STATE $C_i$ calculates $\bold{MOD}({g}_i(\boldsymbol{\theta}^t))=\vec{\nu}_i^t$;
 \STATE $C_i$ finds $\Phi_i^t=\sum_{j \notin \mathcal{G}(C_i)} \phi_{ij}^{t}$; \label{alg:prob}
 \IF{$i \leq l$}
 \STATE $C_i$ masks the content of $\vec{\nu}_i^t$ by $\Phi_i^t$, i.e., it calculates $\mathscr{E}^{\oplus}(\vec{\nu}_i^t)=\vec{\nu}_i^t \oplus \Phi_i^t$ ;
 \STATE $C_i$ sends $\mathscr{E}^{\oplus}(\vec{\nu}_i^t)$ to the PS;
 \ELSE
 \STATE $C_i$ masks the content of $\vec{\nu}_i^t$ by $\Phi_i^t$, i.e., it calculates $\mathscr{E}^{\ominus}(\vec{\nu}_i^t)=\vec{\nu}_i^t \ominus \Phi_i^t$ ;
 \STATE $C_i$ sends $\mathscr{E}^{\ominus}(\vec{\nu}_i^t)$ to the PS;
 \ENDIF
 \ENDFOR
 \STATE PS finds $\sum_{i=1}^S \mathscr{E}(\vec{\nu}_i^t)$;
 \ENDIF
 \end{algorithmic} 
 \end{algorithm}
 
\begin{theorem} \label{Th:decode}
PS can decode the aggregated data.
\end{theorem}
\begin{proof}
Consider two clients $C_i$ and $C_j$. Then, during the coherence time of the channel, the channel reciprocity implies that $\phi_{ij}=\phi_{ji}$ \cite{goldsmith2005wireless}, i.e., the transmitted signals in the links $C_i\rightarrow C_j$ and $C_j\rightarrow C_i$ will experience almost the same fading in the phase.

What aggregates at the PS is 
\begin{flalign}
\bold{Agg}^t&=\sum_{i \in \mathscr{A}} \mathscr{E}^{\oplus}(\vec{\nu}_i^t)+\sum_{i \in \mathscr{B}}\mathscr{E}^{\ominus}(\vec{\nu}_i^t) \label{AGG}
\\
&=\sum_{i \in \mathscr{A}} \vec{\nu}_i^t \oplus \Phi_i^t +\sum_{i \in \mathscr{B}} \vec{\nu}_i^t \ominus \Phi_i^t&
\\
&=\sum_{i \in \mathscr{A},\mathscr{B}}\vec{\nu}_i^t+\sum_{i \in \mathscr{A}}\sum_{j \notin \mathscr{A}} \phi_{ij}^{t} \ominus \sum_{i \in \mathscr{B}}\sum_{j \notin \mathscr{B}} \phi_{ij}^{t}&
\\
&=\sum_{i \in \mathscr{A},\mathscr{B}}\vec{\nu}_i^t+\sum_{i \in \mathscr{A}}\sum_{j \in \mathscr{B}} \phi_{ij}^{t} \ominus \sum_{i \in \mathscr{B}}\sum_{j \in \mathscr{A}} \phi_{ij}^{t}&
\\
&=\sum_{i \in \mathscr{A},\mathscr{B}}\vec{\nu}_i^t+\sum_{i \in \mathscr{A}}\sum_{j \in \mathscr{B}} \phi_{ij}^{t} \ominus \sum_{i \in \mathscr{B}}\sum_{j \in \mathscr{A}} \phi_{ij}^{t}&
\\
&=\sum_{i \in \mathscr{A},\mathscr{B}}\vec{\nu}_i^t+\sum_{i \in \mathscr{A}}\sum_{j \in \mathscr{B}} \phi_{ij}^{t} \ominus \sum_{i \in \mathscr{A}}\sum_{j \in \mathscr{B}} \phi_{ij}^{t}&
\\ 
&=\sum_{i \in \mathscr{A},\mathscr{B}}\vec{\nu}_i^t=\sum_{i=1}^S\vec{\nu}_i^t& \label{AGG_end}.
\end{flalign}
Therefore, PS found the average of updated parameters.
\end{proof}
\begin{theorem} \label{Th:security}
  Algorithm 1 satisfies the privacy requirement for FL.
\end{theorem}
\begin{proof}
First, we show that each client is oblivious to the transmitted data of the other clients. First note that based on Theorem 1, the value $\Phi_i^t$ for all $1 \leq i \leq S$ is uniformly distributed in $[0,2\pi)$. On the other

Without loss of generality, we consider a client in group $\mathscr{A}$. Then 

Second, we show that the PS is oblivious to the data transmitted by all clients. 
\end{proof}


Handling client dropouts: 
When the number of clients is large, the probability that some clients drop out during the training process is high; because a client may become temporarily unavailable due to some reasons such as broken network connections. In this case, PS cannot find decode the aggregated model, because the summation of masked phases will not be equal to zero. In the following, Algorithm 1 is slightly modified so that the PS can accomplish aggregation in case of user drop-out.  

When client $C_i$ drops out, a simple remedy is that the PS asks the remaining clients to send him the phases that contributes to make $\Phi_i^t$. To this end, if $C_i \in \mathscr{A}$, PS would ask the clients in $\mathscr{B}$ to share the channel phases between themselves and $C_i$, and therefore PS would find $\Phi_i^t$ by summing up these phases. Finally, PS subtracts $\Phi_i^t$ from the aggregated model. 

However, there is still a problem associated with this remedy. To shed more light, consider a scenario where $C_i$ transmits his data with some delay to PS. Then, PS presumes that $C_i$ has dropped, and therefore asks the clients in the other group to reveal their contributions in making $\Phi_i^t$. However, right after recovering $\Phi_i^t$, PS will receive the delayed data from $C_i$. Now, the PS is capable of unmasking the content of constellation points transmitted by $C_i$ by de-rotating his constellation points by the secret phase $\Phi_i^t$. Therefore, 
the algorithm requires rectification to handle the dropped users. 

To this aim, each client masks the content of its gradient update by two random phases. The first phase, which is called private random phase hereafter, is a random variable uniformly distributed in $[0,2\pi)$, denoted by $u_i$; and the second phase is $\Phi_i^t$ obtained as discussed in Algorithm 1. A client $C_i \in \mathscr{A}$ mask the content of $\vec{\nu}_i^t$ as follows
\begin{align}
\mathscr{E'}^{\oplus}(\vec{\nu}_i^t)=\vec{\nu}_i^t \oplus u_i \oplus \Phi_i^t,
\end{align}
and similarly, a client $C_j \in \mathscr{B}$ masks the content of $\vec{\nu}_i^t$ as
\begin{align}
\mathscr{E'}^{\ominus}(\vec{\nu}_i^t)=\vec{\nu}_i^t \oplus u_i \ominus \Phi_i^t,
\end{align}
To find the aggregate of
the user models, the PS asks for either (i) the shares of public random phase  $\Phi_i^t$ belonging to the dropped clients, or (i) secret random phase $u_i$ associated with the surviving clients (Note that PS does not ask for both $\Phi_i^t$ and $u_i$, since in that case it can unmask the constellations' contents).
Once the server collects this information, it performs the following calculation over the aggregated model
\begin{align}
\bold{Agg}^t &=\sum_{i \in \mathscr{A}_s} \mathscr{E}^{\oplus}(\vec{\nu}_i^t)+\sum_{i \in \mathscr{B}_s}\mathscr{E}^{\ominus}(\vec{\nu}_i^t)
\\
&-\sum_{i \in \mathscr{A}_d} \Phi_i^t + \sum_{i \in \mathscr{B}_d} \Phi_i^t-\sum_{i \in \mathscr{A}_s,\mathscr{B}_s} u_i \nonumber
\\
&=\sum_{i \in \mathscr{A}_s,\mathscr{B}_s}\vec{\nu}_i^t \label{eq:AGG_mod},
\end{align}
where $\mathscr{A}_{s(d)}$ and $\mathscr{B}_{s(d)}$ denoted the surviving (dropped) clients in group $\mathscr{A}$ and $\mathscr{B}$, respectively. Note that \eqref{eq:AGG_mod} is obtained from the same calculations as in \eqref{AGG}-\eqref{AGG_end}. Thus, PS found the aggregated model. Algorithm 2 illustrates the steps systematically. 

\begin{algorithm} \label{alg:02}
 \caption{Two-group structure with secret/public phases}
 \begin{algorithmic}[1]
 \renewcommand{\algorithmicrequire}{\textbf{Input:}}
 \renewcommand{\algorithmicensure}{\textbf{Output:}}
 \REQUIRE The number of clients, namely $S$
 \STATE Randomly divide the clients into two groups, namely $\mathscr{A}$ and $\mathscr{B}$ \label{alg:grouping};
 \IF{there is a group with less that two clients}
 \STATE{Return to Step \ref{alg:grouping}}
 \ELSE
 \STATE Label the clients in the first group by $\{1,2,...,l\}$, and those in the second group by $\{l+1,l+2,...,S\}$;
 \FOR{$i$ = 1,...,$S$ in parallel do}
 \STATE $C_i$ calculates $\bold{MOD}({g}_i(\boldsymbol{\theta}^t))=\vec{\nu}_i^t$;
 \STATE $C_i$ finds $\Phi_i^t=\sum_{j \notin \mathcal{G}(C_i)} \phi_{ij}^{t}$; \label{alg:prob}
 \IF{$i \leq l$}
 \STATE $C_i$ masks the content of $\vec{\nu}_i^t$ by $\Phi_i^t$, i.e., it calculates $\mathscr{E}^{\oplus}(\vec{\nu}_i^t)=\vec{\nu}_i^t \oplus \Phi_i^t$ ;
 \STATE $C_i$ sends $\mathscr{E}^{\oplus}(\vec{\nu}_i^t)$ to the PS;
 \ELSE
 \STATE $C_i$ masks the content of $\vec{\nu}_i^t$ by $\Phi_i^t$, i.e., it calculates $\mathscr{E}^{\ominus}(\vec{\nu}_i^t)=\vec{\nu}_i^t \ominus \Phi_i^t$ ;
 \STATE $C_i$ sends $\mathscr{E}^{\ominus}(\vec{\nu}_i^t)$ to the PS;
 \ENDIF
 \ENDFOR
 \STATE PS finds $\sum_{i=1}^S \mathscr{E}(\vec{\nu}_i^t)$;
 \ENDIF
 \end{algorithmic} 
 \end{algorithm}

Algorithm 2 satisfies the Conditions I and II, and furthermore can handle the dropped users. Nevertheless, the number of communications between clients needed to establish the mask phases is $O((N/2)^2)$. Furthermore, if $C_i$ drops, PS needs to establish on the average $O(N/2)$ communications with the clients in the other group to reconstruct $C_i$'s random mask. This communication overhead severely limits the network size for real-world applications. In the following Section, Algorithm 2 is rectified so that its communication overhead is significantly decreased. 

\section{sub-groups}
The clients are divided into $K$ groups, namely $\{\mathscr{A}_1,\mathscr{A}_2,...,\mathscr{A}_K\}$, such that the number of clients in each group is $2L$, and thus $N=(K)(2L)$ (if $N=K2L+r$ then we make the last group with $2L+r$ clients instead. This does not make any changes to the proposed algorithm, specifically when the number of clients $N$ is large enough). In the next step, the clients in each group are divided into two subgroups with each having $L$ clients. Denote by $\mathscr{A}^\oplus_i$ and $\mathscr{A}^\ominus_i$ the two subgroups in group $\mathscr{A}_i$.
In order for clients to mask their gradient update constellation points, an approach similar to Algorithm 2 is performed in each of the $K$ groups separately. To elucidate, consider the first group $\mathscr{A}_1$. Then a client $C_i \in \mathscr{A}^\oplus_1$ calculates $\Phi_i^t=\sum_{j \in \mathscr{A}^\ominus_1} \phi_{ij}^{t}$, that is the summation of channel phases between $C_i$ and the clients in the other subgroup of $\mathscr{A}_1$. This procedure is depicted in Fig. \ref{?}. Once the secret masks are achieved, the same procedure as Algorithm 2 is carried out by clients and PS.

In each group, the number of communications between clients needed to establish the mask phases is $L^2$, and as there are $K$ groups, the total number of communications becomes $KL^2=\frac{N}{2L}L^2=\frac{NL}{2}$. 
\begin{remark}
On the one hand, if $K$ increases, then communication overhead $\frac{NL}{2}$ decreases (as $N=(K)(2L)$ is constant). On the other hand, in order to maintain the security of the clients, $L$ cannot be less than 2. Furthermore, if $L=2$, then upon dropping out a client, $L$ becomes equal to 1 which infringes the security of the model. Therefore, there is a trade-off between the communication overhead and the ability of the protocol to handle the dropped clients while maintaining the privacy.
\end{remark}

\bibliographystyle{IEEEtran}
\bibliography{refs}

\end{document}